\documentclass[aps,prl,reprint,groupedaddress]{revtex4-1}

\usepackage{amsmath}
\usepackage{amsfonts}
\usepackage[pdftex]{graphicx}
\usepackage[pdftex]{color} 

\begin{document}



\title{Janus spectra in two-dimensional flows}

\author{Chien-Chia Liu}
\author{Rory T.\ Cerbus}
\author{Pinaki Chakraborty}
\email[]{pinaki@oist.jp}

\affiliation{Fluid Mechanics Unit, Okinawa Institute of Science and Technology 
Graduate University, Onna-son, Okinawa, Japan 904-0495}



\date{\today}

\begin{abstract}

In theory, large-scale atmospheric flows, soap-film flows 
and other two-dimensional flows may host
two distinct types of 
turbulent energy 
spectra---in one,
$\alpha$, the 
spectral exponent of velocity 
fluctuations, equals $3$ and the fluctuations are
dissipated at the small scales, and in the other,
$\alpha=5/3$ and the fluctuations are
dissipated at the large scales---but
measurements downstream of obstacles have 
invariably revealed
$\alpha = 3$. 
Here we report experiments on
soap-film flows where
downstream of obstacles there exists 
a sizable interval in which $\alpha$ has transitioned from $3$
to $5/3$ for the
streamwise fluctuations but remains equal to $3$ for the transverse 
fluctuations, as if two mutually independent 
turbulent 
fields of disparate dynamics were concurrently active within the flow. 
    This species of turbulent energy spectra, 
 which we term the Janus spectra,
   has never been observed or predicted theoretically.
Our results
may open up new vistas 
in the study of turbulence and geophysical flows.
\end{abstract}


\maketitle


Turbulence sculpts clouds. 
By examining the ever-changing patterns in rising
cumulus clouds, L.\ F.\ Richardson
postulated 
the concept of an energy cascade: the mean flow supplies 
turbulent kinetic energy to the large-scale fluctuations,
or large eddies, which  
split to engender smaller eddies, which, in turn,
engender even smaller eddies, and so forth \cite{richardson1922}.
This progressive spawning
of smaller eddies cascades the energy from larger to
smaller scales.
The smallest eddies of the cascade dissipate the energy
viscously.

In 1941, A.\ N.\ Kolmogorov casted the
energy cascade in a mathematical form \cite{KOLM41, *batchelor1947kolmogoroff}.
In this celebrated theory---the phenomenological theory 
of turbulence---Kolmogorov introduced the notion 
of local isotropy. Physically, local isotropy is
based on the idea that
as larger eddies spawn smaller eddies, the 
smaller eddies progressively lose any 
sense of orientation. 
While the large eddies (of size $L$)
are anisotropic, the small eddies (of size $l \ll L$)
are isotropic. 
Assuming local isotropy,
Kolmogorov argued that 
the energy is transferred without dissipation
for a range of small eddies,  $L \gg l \gg \eta$, where
$\eta$ is the size of the smallest eddies that effect viscous
dissipation.
These are the eddies of the ``inertial range.''
In the inertial range,  the turbulent energy spectrum, $E(k)$,
takes a self-similar form, 
$E(k) \sim k^{-\alpha}$,
where $k$ is the wavenumber ($k \sim 1/l$) and $\alpha$ 
the ``spectral exponent.''
Invoking dimensional analysis,
the phenomenological theory 
predicts $\alpha = 5/3$ \cite{dim_analy}.

Consider two-dimensional turbulent 
flows. 
In the 1960s, R.\ H.\ Kraichnan, 
C.\ Leith,
and G.\ K.\ Batchelor adapted 
the phenomenological
theory to 2D turbulent flows
\cite{KRAI67, leith1968diffusion, batchelor1969computation}. 
This theory
predicts
two distinct cascades in
the locally-isotropic small scales.
In the ``direct enstrophy cascade,''
enstrophy is transferred 
without dissipation
from larger
to smaller scales, and
in the ``inverse energy cascade,'' energy is
transferred without dissipation
from smaller to larger scales, inverse of
the 3D energy cascade.

The cascades can be identified via 
$E(k)$. In the inertial range, $E(k) \sim k^{-\alpha}$.
Each cascade
pairs with a specific value of $\alpha$.
Instead of $E(k)$, experiments typically measure
a closely related quantity, 
the one-dimensional
turbulent energy spectra \cite{pope2000tf}: 
the streamwise energy spectrum, 
$E_{uu}(k_x)$, and the transverse energy spectrum $E_{vv}(k_x)$;
$u$ is the streamwise velocity fluctuation,
$v$ the transverse velocity fluctuation, 
$x$ the streamwise direction,
and
$k_x$ the streamwise wavenumber. In the inertial range
$E_{uu}(k_x) \sim k_x^{-\alpha_u}$ and $E_{vv}(k_x) \sim k_x^{-\alpha_v}$.
Local isotropy 
mandates $\alpha_u = \alpha_v = \alpha$.

\begin{figure}
\hspace{-1.1 cm}
\includegraphics[scale = 0.57]{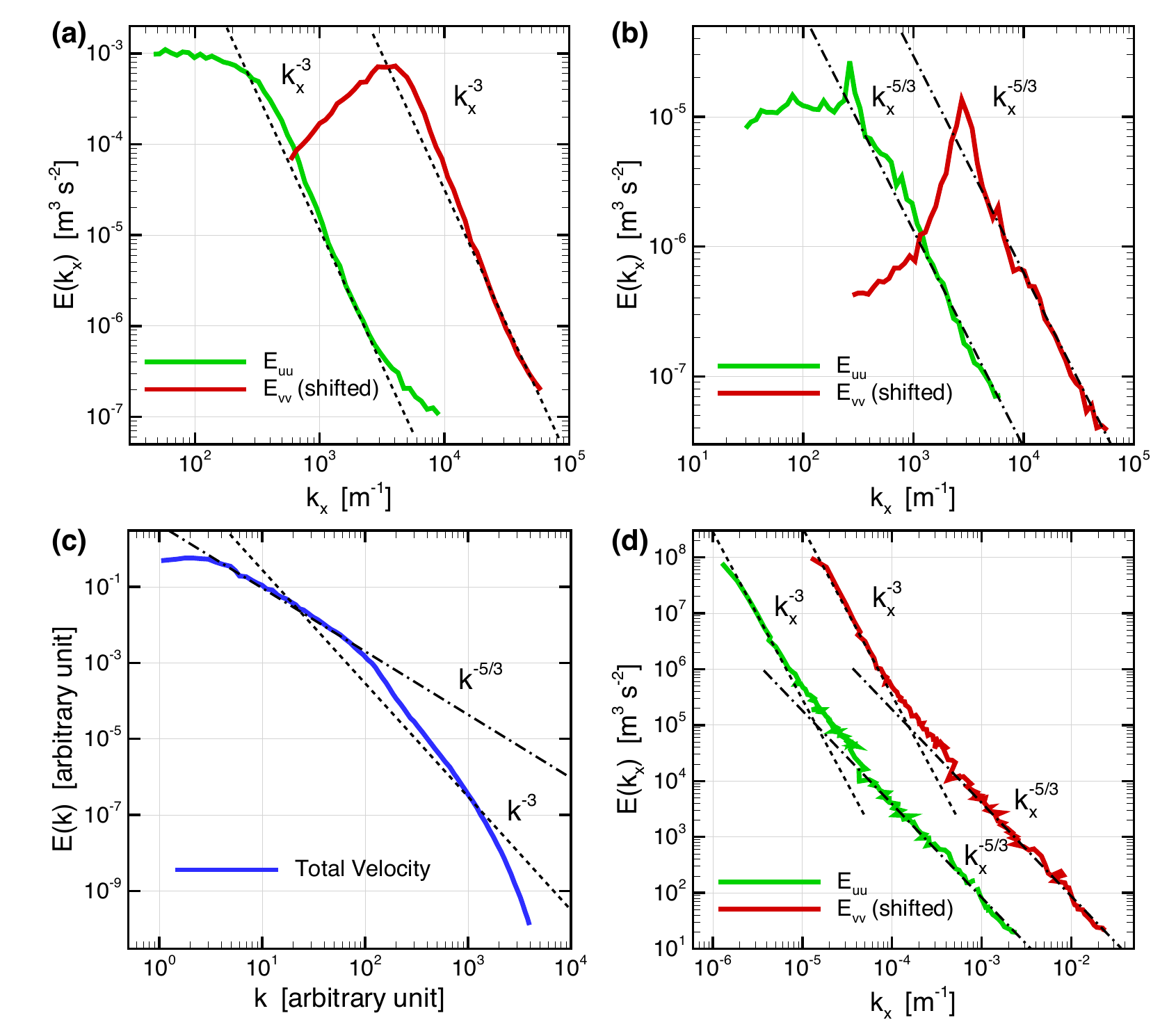} 
\vspace{-0.5 cm}
\caption{
Log-log plots of typical turbulent energy spectra 
in  2D turbulence:
(a) Direct enstrophy cascade 
(experimental setup 
of \cite{Kellay2012, *samanta2014scaling});
 (b) Inverse energy  cascade 
(experimental setup 
of \cite{Kellay2012, *samanta2014scaling});
(c) Double cascade 
     (from \cite{boffetta2010}; 
$\alpha = \alpha_u = \alpha_v$ is implicit in this plot;
$\alpha > 3$ in the 
span of the direct enstrophy cascade is attributed to finite-size
effects); 
(d) Atmospheric cascade (from \cite{nastrom1984}).
Plots of $E_{vv}(k_x)$ (red) are shifted 
from those of $E_{uu}(k_x)$ (green) for clarity.
In all cases
$\alpha_u \approx \alpha_v$.
}
\label{canonical}
\end{figure}

Invoking dimensional analysis, 
the phenomenological theory of 2D turbulence 
predicts $\alpha = 3$ for the 
direct enstrophy cascade
and $\alpha = 5/3$ for the inverse energy cascade
\cite{dim_analy}.
In Fig.~\ref{canonical}a and \ref{canonical}b we show plots of
typical experimental data exhibiting
the direct enstrophy cascade and  
 the inverse energy cascade,
respectively. These two canonical
cascades of 2D turbulence combine to
engender two other well-known cascades.
In the ``double cascade''  (Fig.~\ref{canonical}c),
$\alpha = 5/3$ 
 at low $k$ (inverse energy cascade) and 
$\alpha = 3$ at high $k$ (direct enstrophy cascade).
Large-scale atmospheric flows exhibit a   
    transposed variant of the double cascade 
       (Fig.~\ref{canonical}d).

In all the cases
shown in
Fig.~\ref{canonical},
we note, in accord with local isotropy,
$\alpha_u \approx \alpha_v \approx 3$
or $\alpha_u \approx \alpha_v \approx 5/3$.
By contrast, here 
we report experiments on turbulent soap-film 
 flows in which local isotropy is manifestly
violated;
over a sizable interval of space and over a shared 
span of 
wavenumbers, we find
$\alpha_u \approx 5/3$ (corresponding to the
inverse energy cascade)
and $\alpha_v \approx 3$
(corresponding to the direct enstrophy cascade).
We term 
this ``two-faced'' turbulent energy spectra,
the Janus spectra
(after Janus, the two-faced Roman 
deity).
To our knowledge, any report of such a species of turbulent energy spectra
is unprecedented in 2D turbulent flows \cite{direction_dependent_alpha}. 

Let us briefly consider, for context,
the well-studied case
of decaying 2D turbulence without a mean 
flow \cite{mcwilliams1984, boffetta2012}.
Here the turbulence decays in time. Initially the flow
exhibits
the direct enstrophy cascade, $\alpha_u \approx \alpha_v \approx 3$.
Later the flow 
evolves to 
markedly steeper inertial-range 
energy spectra,
$\alpha_u \approx \alpha_v \sim 4$--$5$.
Local isotropy prevails thoughout the evolution of the flow.

\begin{figure*}
\begin{center}
\includegraphics[scale = 0.62]{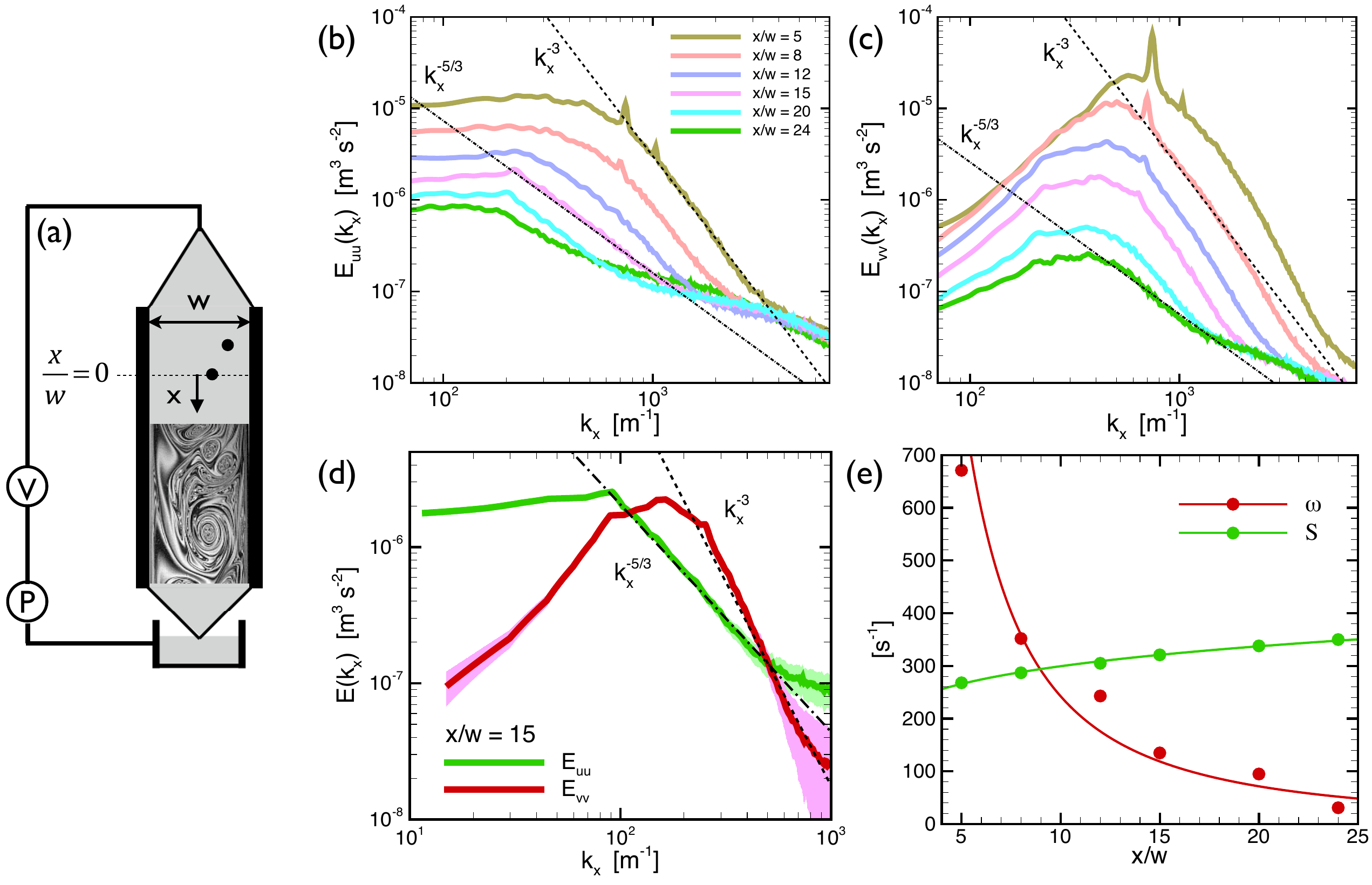} 
\end{center}
\vspace{-0.5 cm}
\caption{Route to the Janus spectra in turbulence
induced by two rods
(also see Fig.~S-2
\cite{supp_mat}).
(a) Schematic of the experimental setup \cite{rutgers2001conducting}.
The $\approx 5 \mu$m 
thick falling soap
film hangs vertically from two steel blades that are
1.6 m long and
w = 12 mm apart from each other.
The soap film
is sustained by recirculating a Newtonian
soap--water solution (1\% commercial detergent in water)
via a pump P and through a valve V (to control the flow rate).
The soap film is pierced with two rods 
(of diameter 1 mm, placed
2 mm and 4 mm transversely away from the centerline, 
and 25 mm vertically staggered from each other).
The mean velocity at the centerline spans 1.60--2.10 m/s.
(b, c) Evolution of $E_{uu}(k_x)$ and $E_{vv}(k_x)$ with
downstream distance. 
(d) Janus spectra at $x/w = 15$.
The shaded regions in
$E_{uu}(k_x)$ and $E_{vv}(k_x)$ represent error bars
(see section S-2 \cite{supp_mat}). 
Note that in the inertial region
the errors are negligible. 
(e) Evolution of turbulent vorticity ($\omega$) and 
mean shear ($S$) with
downstream distance (see section S-3 \cite{supp_mat}).
The best-fit lines are to guide the eye.
}
\label{two_rods}
\end{figure*}

We conduct experiments in a soap-film channel, 
a well-known setup
for studying quasi-2D turbulent flows (Fig.~\ref{two_rods}a).
To induce turbulence in the soap film, we pierce it with
two rods non-symmetrically about the centerline.
(This setup emulates an atmospheric flow; see Fig.S-1 \cite{supp_mat}.)
The rods shed eddies as the film squeezes past them.
The eddies render the flow turbulent, which
decays downstream of the rods. 
A decaying 2D turbulent flow is a canonical case
for the direct enstrophy cascade \cite{batchelor1969computation}.
Indeed, numerous experiments of soap films flowing downstream of obstacles
attest to $\alpha_u \approx \alpha_v \approx 3$
(see, e.g., \cite{kellay2002, *martin1998, *rivera2014}).

To interrogate the turbulent flow 
we turn to
Laser Doppler Velocimetry (LDV).
We use LDV
to measure the time-series of 
$u$ and $v$
along the centerline of the channel,
downstream of the rods and
well upstream ($> 15 w$)
of the Marangoni shock \cite{tran2009marangoni}.
Using Taylor's hypothesis 
\cite{taylor1938spectrum, *belmonte2000experimental}
we transform this time-series data to spatial data
in the streamwise direction, from which
we compute $E_{uu}(k_x)$ and $E_{vv}(k_x)$.
As can be seen from
the decrease in the amplitude of the energy
spectra (Fig.~\ref{two_rods}b,c), or
from the decay of the turbulent vorticity (Fig.~\ref{two_rods}e),
the turbulence decays downstream of the rods.
Not far downstream 
($x/w \lesssim 10$), 
$\alpha_{u} \approx \alpha_{v} \approx 3$ (Fig.~\ref{two_rods}b,c),
in accord with previous experiments (and also consistent with
the 
initial phase of decaying 2D turbulence without a mean flow). 
Note that the mechanisms that sustain the direct enstrophy cascade
entail strong eddy--eddy
interactions \cite{chen2003, boffetta2012}. 
Therefore, experiments targeted at
the direct enstrophy cascade focus on the region 
where the eddies have not decayed significantly---the
region near the obstacles.
Here we depart from these experiments 
in an unremarkable 
way: we continue to measure the energy spectra farther downstream.
And yet, this unveils a series of remarkable features.
 
Extrapolating from the later phase 
of decaying 2D turbulence without a mean flow, 
downstream of the direct enstrophy cascade
we expect 
steeper energy spectra with $\alpha_u \approx \alpha_v > 3$.
The mean flow in the experiment, however, 
effects a strikingly different 
fate.
(We return to the role of the mean flow later.)
For $x/w \gtrsim 12$, 
$\alpha_u$ begins to
decrease but $\alpha_v$ remains $\approx 3$
(Fig.~\ref{two_rods}b,c).
The regime of local isotropy is broken.
Farther downstream
($15 \lesssim x/w \lesssim 20$) 
$\alpha_u$
reaches a plateau $\approx 5/3$, suggesting
an unexpected 
transition to the inverse energy cascade for $E_{uu}(k_x)$.
In this region, remarkably, $\alpha_v$ is still $\approx 3$,
suggesting the persistence of the
direct enstrophy cascade for $E_{vv}(k_x)$.
That is, over a sizable interval
($\approx 5w$) and over a shared 
span of wavenumbers,
$\alpha_u \approx 5/3$ but concurrently
 $\alpha_v \approx 3$ (Fig.~\ref{two_rods}d). 
This is the domain of the Janus spectra.
(Even farther downstream, both $\alpha_u$
and $\alpha_v$ decrease monotonically.)
We note, for contrast, that 
previous experiments in soap-film channels have reported the
inverse energy cascade for turbulence forced by rough blades
(\cite{Kellay2012, *samanta2014scaling}; 
Fig.\ref{canonical}b). This roughness-induced
turbulent flow exhibits $\alpha_u \approx \alpha_v \approx 5/3$,
in accord with local isotropy. 
Our experiments, on the other hand, have smooth blades and
we find $\alpha_u \approx 5/3$,
but without local isotropy, $\alpha_u \neq \alpha_v$.

\begin{figure*}
\begin{center}
\includegraphics[scale = 0.5]{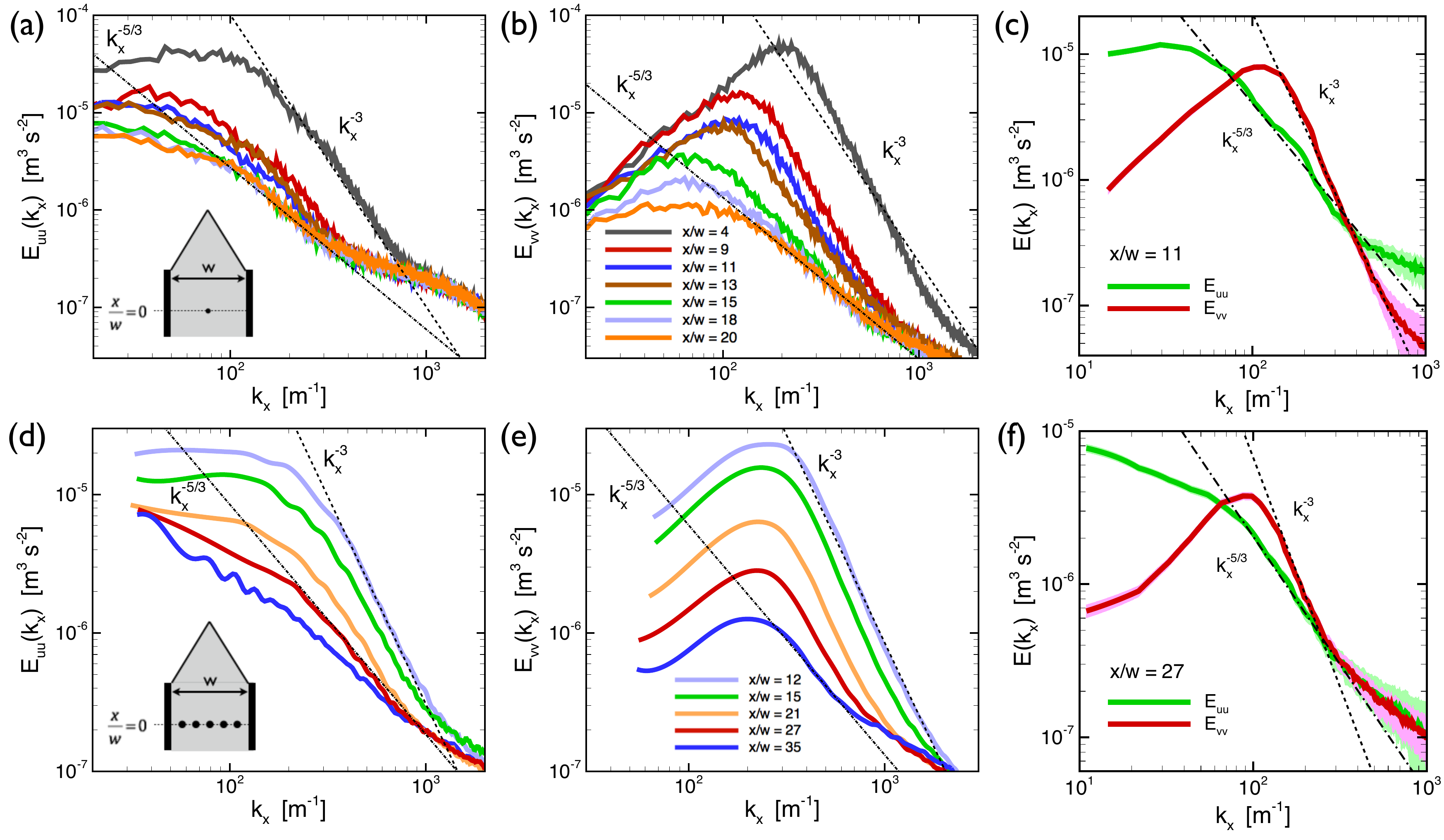} 
\end{center}
\vspace{-0.5 cm}
\caption{
Route to the Janus spectra in turbulence
induced by one rod (a--c) and comb (d--f);
(also see Figs.~S-3
and S-4 \cite{supp_mat}).
One rod: The soap film (w = 23 mm)
is pierced at the centerline with a rod of diameter 0.5 mm;
the mean velocity at the centerline spans 2.10--2.40 m/s.
Comb: the soap film (w = 17 mm)
is pierced symmetrically about the centerline with a 
comb (five 
rods of diameter 1 mm, spaced 3 mm apart 
from each other);
the mean velocity at the centerline spans 
2.07--2.60 m/s.
(a,b; d,e) The evolution of $E_{uu}(k_x)$ and $E_{vv}(k_x)$ with
downstream distance is similar to 
the case of the two rods (Fig.~\ref{two_rods}).
We show the Janus spectra in (c) for one
rod ($x/w = 11$) and in (f) for comb ($x/w = 27$);
the errors (the shaded regions) 
are negligible in the inertial range.
Note that even though the wavenumber span of
 the inertial region of $E_{vv}(k_x)$ in (f)
 is relatively small, we can infer
 $\alpha_v \approx 3$ by comparing with the energy spectra
 measured upstream (e).
 Upstream, $\alpha_v$ is the same, but the span of the inertial
 region is considerably
 broader. (Similar considerations also apply to the
 $E_{vv}(k_x)$ for two rods and one rod.)
}
\label{rod_comb}
\end{figure*}

Because the dynamics of decaying 2D turbulence is
sensitive to the mode of forcing turbulence \cite{boffetta2012},
we test
if the Janus spectra is
unique to
the non-symmetric forcing by the two rods.
To that end
we conduct experiments with two
different obstacles:
one rod and
comb. (A comb, 
the standard choice of
forcing in soap-film experiments,
is a row of rods.)
We place the obstacle symmetrically about the
centerline of the soap-film channel.
Proceeding downstream from the obstacle, 
we first see $\alpha_u \approx \alpha_v \approx 3$;
farther downstream we note that
$\alpha_u$ transitions to 
a plateau $\approx 5/3$ 
but $\alpha_v$ remains $\approx 3$---the Janus spectra
(Fig.~\ref{rod_comb}). 
We conclude that the Janus
spectra is a robust phenomenon, unaffected by the symmetry
(or lack thereof) of the forcing.

The existence of the Janus spectra suggests 
that the phenomenological theory may be extended to flows
without local isotropy. 
Interestingly, in 
the Janus spectra, it is not
simply that $\alpha_u \neq \alpha_v$, 
where $\alpha_u$ and $\alpha_v$ assume values 
whose interpretation necessitates
a new theoretical framework.
Instead, we find 
$\alpha_u \approx 5/3$ and $\alpha_v \approx 3$,
the same spectral exponents as those predicted by
the phenomenological theory for locally-isotropic 
2D turbulent flows. 
Based on our empirical results, we postulate a simple
generalization of the phenomenological theory for
the Janus spectra:
the $u$ component transfers energy
without dissipation and the
$v$ component transfers enstrophy
without dissipation. Consequently,
dimensional analysis 
\cite{dim_analy}
yields $\alpha_u = 5/3$ and $\alpha_v = 3$.

\begin{figure}
\begin{center}
\includegraphics[scale = 0.4]{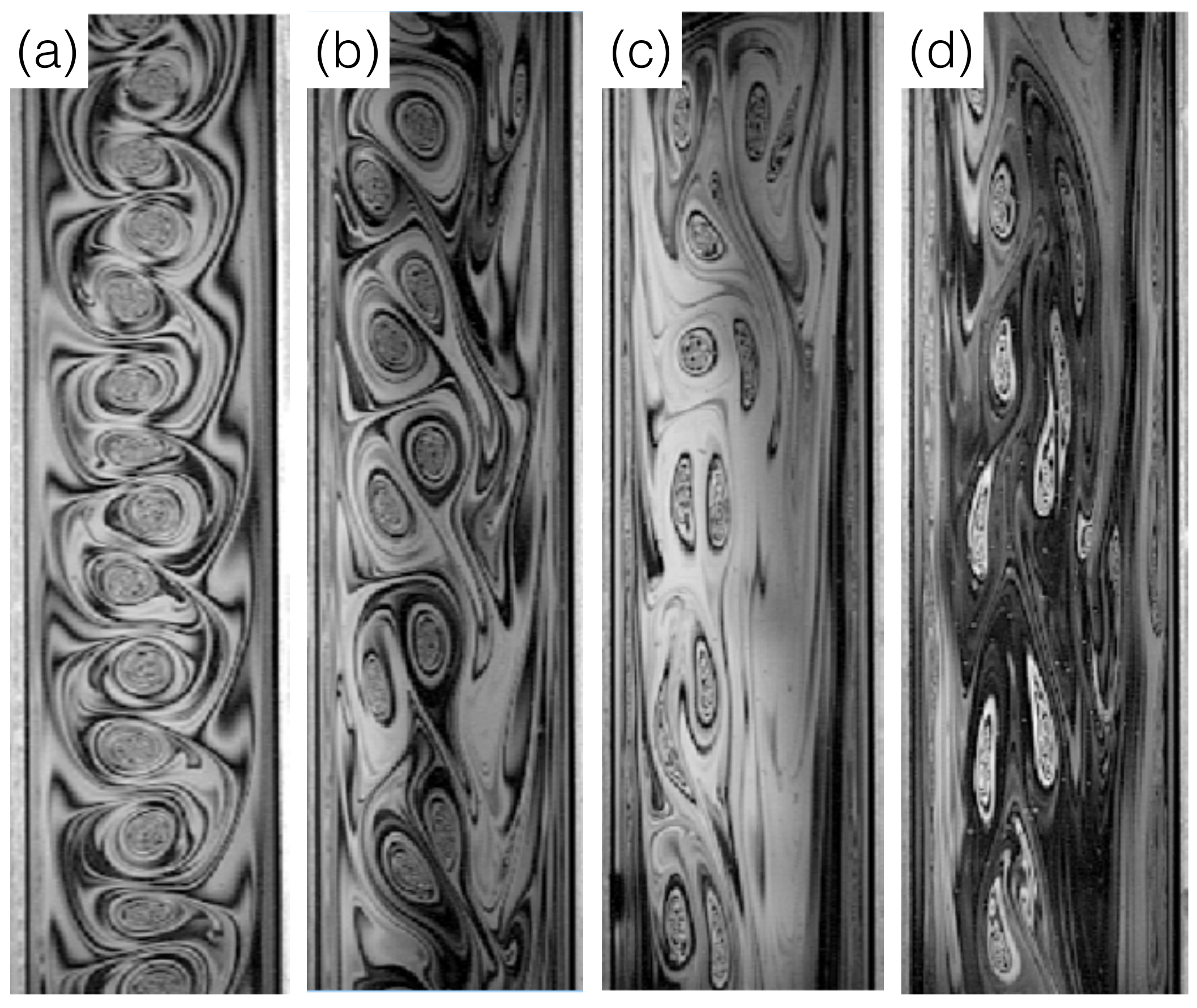} 
\end{center}
\vspace{-0.5 cm}
\caption{Turbulent eddies distorted by the sheared mean flow.
The soap film (w = 22 mm)
is pierced at the centerline with a rod of diameter 0.5 mm.
The mean flow in each panel is from top to bottom; 
from left to right, the panels
document the downstream evolution of the flow: 
$x/w \approx$ 1--4 (a), 
6--9 (b), 25--28 (c) and 29--32 (d). 
}
\label{fringes}
\end{figure}

To seek the physical mechanisms that underlie the Janus
spectra, we turn to flow visualization.
We illuminate the soap film with
monochromatic light 
(wavelength = $633$ nm).
The resultant interference fringes
render the turbulent eddies visible (Fig.~\ref{fringes};
Fig.~S-5 \cite{supp_mat}). 
The shapes of the eddies bear witness to the influence of
the sheared mean flow in the soap-film channel. 
Just downstream of the obstacle, the eddies are 
isotropic (Fig.~\ref{fringes}a,b). Farther
downstream, however, the eddies become progressively 
anisotropic
(Fig.~\ref{fringes}c,d).
This transition from isotropy to anisotropy---and the
accompanying
transition from the
direct enstrophy cascade to the Janus spectra---can be
understood by comparing the magnitudes
of the turbulent vorticity and the mean shear: 

As the flow evolves downstream,
the turbulent vorticity decays but the mean shear
remains roughly the same (see, e.g., Fig.~\ref{two_rods}e).
Upstream, the turbulent eddies prevail. 
Being relatively unaffected by the mean shear, the 
locally-isotropic flow
exhibits the direct enstrophy cascade (similar to
the 
case of decaying 2D turbulence without a mean flow).

Farther downstream, the turbulent eddies weaken.
The weak eddies are distorted by the mean shear
\cite{kida1981, *marcus1990, *toh1991, *nycander1995, *cummins2010}, as 
is evident in Fig.~\ref{fringes}c,d. 
This effects 
anisotropy across the scales, large and small.
To get to a heuristic picture of the mechanisms that
engender the Janus spectra,
we proceed by using the
shapes of these anisotropic eddies as a guide.
Note that the anisotropic eddies of Fig.~\ref{fringes}c,d 
are preferentially elongated along the streamwise direction
(also see \cite{kida1981, *marcus1990, *toh1991, *nycander1995, *cummins2010}), 
thereby
transferring energy for the $u$ component 
to larger scales (section S-5
\cite{supp_mat}; \cite{lift-up}).
We postulate that this energy transfer initiates 
the inverse energy cascade
for the $u$ component.
The $v$ component of these anisotropic eddies
exhibits different dynamics.
Note that the enstrophy flux 
is modulated by the vorticity gradient \cite{chen2003},
which is oriented
along $v$ for the streamwise-elongated
eddies. 
We speculate that the $v$-oriented
vorticity gradient sustains the transfer of
enstrophy for the $v$ component.
These velocity-component dependent transfers of energy and
of enstrophy manifest as the Janus spectra.

While our discussion offers hints about the
role of mean shear 
\cite{biferale2002anomalous,*biferale2005anisotropy, shear_comment},
 the interaction between mean flow and
turbulent vorticity  remains
poorly understood.
This interaction dictates the fate
not only of soap-film flows but also of
decaying flows in the atmosphere and the ocean, for instance,
the waning phase of tropical cyclones. In these
quasi-2D decaying turbulent flows 
that are inextricably embedded in
sheared mean flows,
anisotropy is but the norm.
Our study suggests that the streamwise
and transverse fluctuations in these flows
partake in mutually-independent 
dynamics,
and that the tools of the phenomenological theory can be
invoked to predict their behavior.
That a theory built assuming local isotropy
can still elucidate a manifestly anisotropic turbulent flow
strikes us as an extraordinary testament to its generality. 
In closing, we submit that 
the study of decaying 2D turbulence
 embedded in a sheared mean  flow
is replete with unexpected insights,
with implications on 
a broad range of problems,
 from the understanding of turbulent cascades to 
 the forecasting of
 large-scale weather systems.

\begin{acknowledgments}
We thank Gustavo Gioia for helpful discussions,
and the referees for constructive comments.
This work was supported by
the Okinawa Institute of Science and 
Technology Graduate University.
\end{acknowledgments}
 

%

\end{document}